\documentstyle[12pt,epsf]{article}
\def\Z{\hbox{{\sf Z}\kern-0.4em {\sf Z}}}
\def\nhat{{\hat n}}

\def\mhat{{\hat m}}

\date{}
\begin{document}

\title{Chern-Simons Correlations on $(2+1)D$ Lattice} 

\author{{\large L. A. Abramyan, 
A. P. Protogenov and V. A. Verbus$\,^{\ddagger}$}}
\maketitle
\begin{center}
{\footnotesize {\em 
Institute of Applied Physics, Russian Academy of Sciences,\\
46 Ul'yanov Street, Nizhny Novgorod 603600 \\
$^{\ddagger}$Institute for Physics of Microstructures, Russian Academy of Sciences,\\
46 Ul'yanov Street, GSP-105, Nizhny Novgorod 603600}}
\end{center}

{\small\parbox{12cm}{\small
We have computed the contribution of zero modes
to the value of the number of particles
in the model of discrete $(2+1)$-dimensional
nonlinear Schr\"odinger equation.
It is shown for the first time
that in the region of small values of
the Chern-Simons coefficient $k$ there exists
a universal attraction between
field configurations. For $k=2$ this phenomenon
may be a dynamic origin of the semion pairing in high temperature
superconducting state of planar systems.  \\[5pt]
}}\\\\\\

The cooperative behavior based on universal topological
features of planar systems is the subject of a number
of papers \cite{Wil,Prot}. The topological
properties of $(2+1)$-dimensional systems are described by a
Chern-Simons term in the Lagrangian of the model.
This term displays the phenomenon of the chiral 
invariance violation in such systems. 
The magnetic Chern-Simons field 
leads usually \cite{Jack,Hon,Wei,Bar,Ber,Pas} to the effective 
repulsion between field configurations. 
This property was described in detail in 
Refs. \cite{Ab2,Ab3} where the structure of zero modes in 
continuous \cite{Ab2} and spacial discrete \cite{Ab3} models 
of $(2+1)D$ nonlinear Schr\"odinger equation was studied. 
The competition of the basic nonlinearity corresponding to 
attraction, diffraction and additional nonlinearity, 
describing the repulsion caused by Chern-Simons interaction, 
resulted in the increase of the critical value of the 
number of particles $N$ in the region of small values of coefficient $k$. 
The increase of $N$ represents the existence of additional 
repulsion.

We argue in this paper that on the $(2+1)D$ lattice
there is universal attraction due to
Chern-Simons correlations between field configurations
and show the conditions when this phenomenon takes place.
We shall show that consideration of complete contribution
of the statistical Chern-Simons fields
in the form of the holonomies causes the additional
attraction (to the bare one) at small numbers $k$ of links. 
The necessary condition for that is naturally the existence of bare
attraction between field configurations.
This result is the first indication that there
exists the universal attraction due to Chern-Simons correlation.
The comparison with the results of our
previous papers \cite{Ab2,Ab3} shows that the
condition for obtaining the attraction is
the inclusion in the consideration the Wilson and Polyakov
exponents on the equal footing. In other words,
we consider the model which takes into account
the conditions of the compactness for the temporary component of the
gauge potential as well as for its spatial component.
The gauge invariance requirements lead immediately
that discrete evolution should be considered (for detail see \cite{Ab3}).
We want to emphasize that the gauged discrete $(2+1)D$
nonlinear Schr\"odinger equation gives us a convenient tool
to display this $(2+1)D$ system phenomenon of additional attraction
which has a general and universal character.

The equation of motion in the model of the gauged discrete $(2+1)D$
nonlinear Schr\"odinger equation has the form
\begin{equation}
({\hat t}_{x} + {\hat t}_{y} + h.c. - 4)\rho_{m,n} =
-2C\rho_{m,n}^{3} - \rho_{m,n}\sin (w_{m,n} - 1)  \, . 
\end{equation}
Here ${\hat t}_{x}=e^{iA_{\mhat,n} + \partial /\partial m}$ 
is the operator of so-called magnetic
translations. The parameter $C=g\;|k|$ in Eq.(1)
contains the coupling constant $g$ of the classical nonlinear Schr\"odinger
equation and Chern-Simons coefficient $k$. Besides the
consideration of the model on $2D$ lattice \cite{Ab3}
we included  the discrete time $t \in \Z$ in the description.
This leads for the stationary states
$\Psi (m,n,t) = \rho_{m,n}e^{it}$ to the existence of
nonlinearity presented by the sine function in Eq.(1).
The multiconnection of the $2D$ manifold has been taken into account
by the gauge field $A^{\mu}(m,n)=(w_{m,n}, A_{\mhat,n}, A_{m,\nhat})$
where
$$
w_{m,n}=\sum\limits_{m',n'}
[(
\Delta_{2}G(m-m', n-n'))
(\rho^{2}_{m',n'}+\rho^{2}_{m',n'+1})A_{m',\nhat'} -
$$
\begin{equation}
(\Delta_{1}G(m-m', n-n'))(\rho^{2}_{m',n'}+\rho^{2}_{m'+1,n'} )
A_{\mhat',n'}]
\end{equation}
is the temporal component of Chern-Simons potential and
\begin{equation}
A_{\mhat,n} = \sum\limits_{m',n'}\Delta_{2}G(m-m',n-n')\rho^{2}_{m',n'}
\end{equation}
is the $x$-component of the vector potential.
The notation $A_{\mhat,n}$ denotes that the components
$A_{x}(m,n)$ are determined on the links connecting the
sites $(m,n), (m+1,n)$. In Eqs.(2),(3)
$\Delta_{1,2}f({\bf r}) \equiv f({\bf r}+
{\bf e}_{1,2}) - f({\bf r})
$ is the gradient on the lattice with coordinates
${\bf r}=(m, n) \in \Z ^{2}$; ${\bf e}_{i}$ is the unit vector. 
In accordance with the rules of the gauge field theory on 
the lattice we assumed that while the gauge field is defined on 
the lattice links, the curl of the field 
$A_{\mu}({\bf r})$ and the density $\rho^{2}$ are defined 
on the sites of the dual lattice. 
Green function on the lattice in Eqs.(2),(3) has the form:
\begin{equation}
G(m-m',n-n') = \int\limits_{-\pi}^{\pi}\,\frac{d^{2}k}{(2\pi )^{2}}
\frac{e^{i\{k_{x}(m-m')+k_{y}(n-n')\}}-1}{4-2\cos k_{x} -2\cos k_{y}} \, .
\end{equation}

The main goal of this paper is to study the dependence of
the critical number of particles
$
N = \sum\limits_{m',n'}\rho^{2}_{m',n'}
$
on the parameter $C=g|k|$
considering the arbitrary large contribution
of the temporal component of the gauge potential as well as
of its spatial component.
To solve the problem
we compute this dependence using the zero modes found for various
values of Chern-Simons coefficient $k$.

We performed the simulation of the problem (1)-(4) on the
lattice with linear size $L \leq 20$ using the method of the stabilizing
Petviashvili multiplier \cite{Pet}. The description of the
block diagram of the method and the details of calculations
may be found in Refs. \cite{Ab2,Ab3}. 
The form of the functions $\rho_{m,n}, A_{\mhat,n}, w_{m,n}$
found numericaly is displayed in Ref.\cite{Ab3}. 
The form of these functions in the present paper 
is qualitatively the same. 
We used the function $\rho_{m,n}$ for the calculation of
the dependence $N(C)$  which is shown in Fig. 1.
Note that the limit $k \to \infty$ is equivalent to the
zero contribution of the gauge fields ${\bf A}_{m,n}, w_{m,n}$,
when in the continuous limit $N_{cr}=11.703$. 

This number of
particles separates the $2D$ collapse regime
at $N > N_{cr}$ and its absence at $N < N_{cr}$. On the lattice
the values $N_{cr}^{lat}$ are always less \cite{Rizh} than the
critical number of particles in the continuous limit even if Chern-Simons
fields are neglected.
Therefore the main problem of interest is
whether $N_{cr}^{lat}(A_{\mu} \neq 0) $ is smaller or greater than
$N_{cr}^{lat}(A_{\mu}=0)$ if the Chern-Simons
gauge fields are taken into account.

\begin{figure}
\epsfxsize =10cm \epsfbox{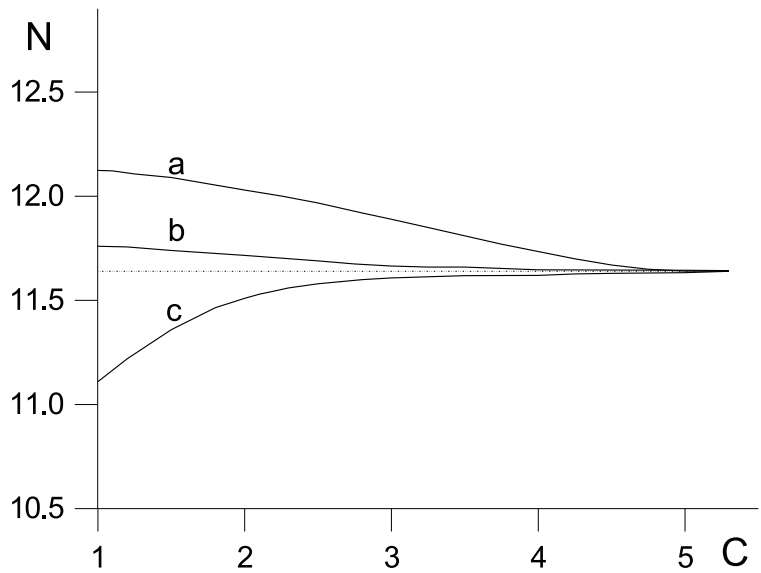}
\caption{
The dependence of the critical number
of the particles on the parameter $C$
for the three types of the
nonlinearity in Eq.(1) (see for details the text).
The dashed line shows the value
$N^{lat}(A_{\mu}=0)=11.605$.
}
\end{figure}

The curve "a" in Fig.1 shows the result of the paper [10]
when we considered only the part $4 - 2\cos A_{x} - 2\cos A_{y}$
of the contribution to Eq.(1) of the spatial gauge field components
and did not consider the discrete time. The case "b" in Fig.1
corresponds to complete consideration of spatial gauge field
contribution to the l.h.s of Eq.(1) with the same properties of the
time as it is above. The dependence $N(C)$ in the case "c" in
Fig.(1) presents the result of computations on the
$(2+1)D$ lattice with consideration of the complete contribution
of all (arbitrary large) gauge field components $A_{\mu}$
taking into account the discrete time.

From dependence $N(C)$ in the case "c" in
Fig.(1) one can infer that the decrease of the critical
number of particles $N$
with decreasing of the parameter $C$
is equivalent to the increase of the attraction
in comparison with the case when we do not take into account
the contribution of the Chern-Simons fields.

It is known \cite{Wil,Prot} that in Chern-Simons systems there
arises the induced angular momentum proportional to $1/k$.
The calculated lines of the equal value of the field
$\rho_{m,n}$ for $C=1$ are shown in Fig. 2.
Here we would like to pay
attention to the extended $s$-symmetry of the field $\rho_{m,n}$
for a small value of the parameter $C=1$ at
great distance from the origin.
We found weak display of this phenomenon (see Fig. 2).
The accuracy of the performed calculations according to our
estimates is several percents.
The comparison of the results on Fig.2 for the different
value of the parameter $C$ shows that the display of
extended $s$-wave symmetry increases with decrease of parameter $C$.

\begin{figure}
\epsfxsize =10cm \epsfbox{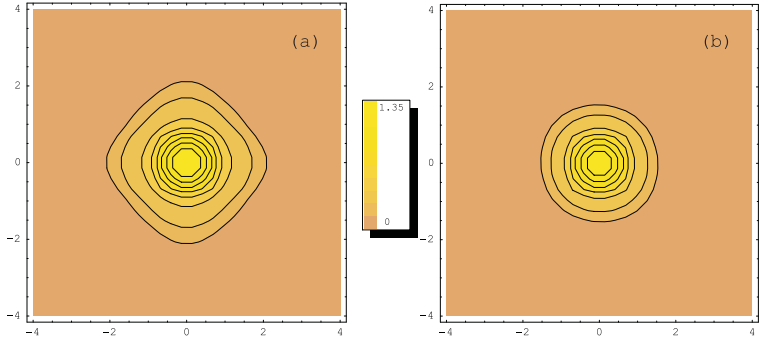}
\caption{
The lines of equal value of the field $\rho_{m,n}$
for $C=1$ (a) and $C=5$ (b).
}
\end{figure}

The origin of the considered phenomena is as follows.
We consider the nonlinear nonlocal dependence of the components
$A_{\mu}(m,n)$ of Chern-Simons gauge field
via the field $\rho_{m,n}$ in its complete form
presented in Eq.(1) for the Laplacian.
In particular, if we extract \cite{Ab3} a part of
this contribution to nonlinearity,
specifically taking into account the compact version
$\rho (4-2\cos A_{x}-2\cos A_{y})$  of the nonlinearity
$\rho(A_{x}^{2}+A_{y}^{2})$ \cite{Ab2}, the rest part in
the discrete Laplacian in the continuous limit has the form
$
\cos A_{x}\frac{\partial^{2}}{\partial x^{2}} +
\cos A_{y}\frac{\partial^{2}}{\partial y^{2}} \, .
$
The decrease of the coefficients in this expression in comparison
with the unity for the value of $A_{m,n}$ in the region of small
coefficients $k$ leads to the decrease of diffraction.
It is seen that the origin of additional attraction due to Chern-Simons
fields is the decrease of the diffraction.
Neglecting this effect we observe \cite{Ab3} only
the repulsion due to Chern-Simons fields. The anisotropy of
this operator is the reason of extended $s$-symmetry of the ground state.

As it was pointed out above the considered phenomenon exists on
the lattice under the condition of the complete taking
into account the gauge field by the holonomies
$e^{iA_{\mu}}$. The physical interpretation is clear:
because the zero component
of gauge potential plays the role of chemical potential
the arbitrary large value of the one corresponds to arbitrary
large value of the energy which we add to the system
adding a particle.

Note that discrete models are
characterized by the features which are absent in the continuous limit.
In particular, the localized states can exist in $(1+1)D$
discrete nonlinear Schr\"odinger equation \cite{Bish}.
In our case the dimensionality of the problem as well
as the discrete character of the space and the time are important.
Using the arguments in inverse order we have to include in the model
discrete space and time on the equal footing in order to
consider the large magnitude of the gauge field in the form of
Wilson exponent as well as Polyakov exponent having in mind
the gauge invariance. Our simulation shows for example that
without the condition of the discretization of the time we can not
obtain the attraction due to Chern-Simons fields.

Finally, we should like to make a general remark. The attraction between
particles in the systems with Chern-Simons interaction was a subject
of extensive studies during the last ten years. The attention was
focused \cite{Kog,Kogpol,Kogkhven} on analyzing the symmetry of the state
($s$-, $p$- or $d$-wave states) with non-zero value of the
superconducting gap in the framework of the perturbation theory
when the parameter $\alpha = 4\pi/|k|$ was small, i.e. in the
limit $|k| \gg 1$. The discovery \cite{Luke} of the time-reversal
symmetry-breaking $p$-wave superconductivity in $Sr_{2}RuO_{4}$
(see also Refs.\cite{Krish,Laugh,Volov}) stimulated the recent
paper \cite{Goryo} where the search for the induced Chern-Simons
term in $P$- and $T$-violating superconductors was performed.

The picture of the Chern-Simons correlations,
described in the present paper, is in some sense beyond the
above-mentioned approaches. We found the attraction due to the
Chern-Simons gauge field in the essentially non-perturbative
region of {\em the small}\/ value of the coefficient $k$ irrespectively of
the symmetry of the ground state. 
The choice of the coefficient $k$ itself determines the 
symmetry of the considered state. In this respect, Fig.2 presents 
only one of the possible symmetry of the ground state characterized 
by the specific value of the parameter $C$.

Let us suppose in the Eq.(1) that $C=0$. We obtain in this case
the model describing the nonlocal interaction
of Chern-Simons vortices. 
This case corresponds in continuous limit to universal 
nonlinearity of $\rho^{5}$ kind (from the point of view of scale 
transformations) and differs from Gross-Pitaevskii model \cite{Pit} 
when the nonlinearity in the equation of motion is a local one and 
proportional to $\rho^{3}$. We plan to study this 
interesting limit in a separate paper.

In conclusion, we studied the dependence of the critical 
particle number on the link numbers of the field 
configurations. Using the model of the discrete $(2+1)D$ nonlinear 
Schr\"odinger equation we found for the first time the existence of 
the attraction due to Chern-Simons fields. 
It was shown that the origin of this phenomenon is the suppression of the 
free propagation by Chern-Simons fields at small link numbers. 
Note that for $g=1$ the semion value $k=2$, which is of topical interest, 
lies inside this region. 
Therefore the found attraction may be the dynamical reason of the 
semion pairing and phase transition to a superconducting state.

We are grateful to S.N. Vlasov, E.A. Kuznetsov, A.G. Litvak,
A.M. Satanin, V.I. Talanov and V.E. Zakharov for very
interesting discussions and useful advices.
The work was supported in part by the RFBR under
Grant No. 98-02-16237.


\newpage

\begin{thebibliography}{99}

\bibitem{Wil} {\em Fractional Statistics and Anyon
Superconductivity},\/ ed. F. Wilczek, World Scientific,
Singapore, 1990.
\bibitem{Prot} A.P. Protogenov, Sov. Phys. Usp.
{\bf 35} (7), 535 (1992).
\bibitem{Jack}
R. Jackiw and S.Y. Pi, Phys. Rev. Lett. {\bf 64}, 2969
(1990); (C) {\bf 66}, 2682 (1991); Phys Rev. D {\bf 42},
3500 (1990); Prog. Theor. Phys. Suppl. {\bf 107}, 1 (1992).
\bibitem{Hon} J. Hong, Y. Kim, and P.Y. Pac,
Phys. Rev. Lett. {\bf 64}, 2230 (1990).
\bibitem{Wei} R. Jackiw and E. Weinberg,
Phys. Rev. Lett. {\bf 64}, 2234 (1990);
R. Jackiw, K. Lee, and E. Weinberg,
Phys. Rev. D {\bf 42}, 3488 (1990).
\bibitem{Bar} I. V. Barashenkov and A.O. Harin,
Phys. Rev. Lett. {\bf 72}, 1575 (1994);
Phys. Rev. D {\bf 52}, 2471 (1995).
\bibitem{Ber} L. Berg\'e, A. de Bouard, and J.C. Saut,
Phys. Rev. Lett. {\bf 74}, 3907 (1995).
\bibitem{Pas} M. Knecht, R. Pasquier, and J.Y. Pasquier,
J. Math. Phys. {\bf 36}, 4181 (1995).
\bibitem{Ab2} L.A. Abramyan, V.I. Berezhiani, and A.P. Protogenov,
Phys. Rev. E {\bf 56}(5), 6026 (1997).
\bibitem{Ab3} L.A. Abramyan, A.P. Protogenov, and V.A. Verbus,
Zh. \'Eksp. Teor. Fiz. {\bf 114}, 747 (1998).
\bibitem{Rizh} E.W. Laedke, K.H. Spatschek, V.K. Mezentsev,
S.L. Musher, I.V. Ryzhenkova, S.K. Turitsyn, Pis'ma v ZhETF
{\bf 62}(8), 652 (1995).
\bibitem{Pet} V.I. Petviashvili, Fizika plazmy {\bf 2}(3), 469
(1976).
\bibitem{Bish} D. Cai, A.R. Bishop, and N. Gr\o nbech-Jensen,
Phys. Rev. Lett. {\bf 72}, 591 (1994).
\bibitem{Kog} Ya.I. Kogan, Pis'ma v ZhETF {\bf 49}, 194 (1989).
\bibitem{Kogpol} Ya.I. Kogan, I.V. Polubin,
Pis'ma v ZhETF {\bf 51}, 496 (1990).
\bibitem{Kogkhven} D.V. Khveshchenko, Ya.I. Kogan,
Pis'ma v ZhETF {\bf 50}, 137 (1989);
Mod. Phys. Lett. {\bf B}4, 95 (1990).
\bibitem{Luke} G.M. Luke {\em et al}., Nature {\bf 394}, 558 (1998).
\bibitem{Krish} K. Krishana {\em et al}., Science {277}, 83 (1997).
\bibitem{Laugh} R.B. Laughlin, Phys. Rev. Lett. {\bf 80}, 5188 (1998).
\bibitem{Volov} G.E. Volovik, Pis'ma v ZhETF {\bf 66}, 492 (1997).
\bibitem{Goryo} J. Goryo and K. Ishikawa, cond-mat/9812412.
\bibitem{Pit} E.M. Lifshits and L.P. Pitaevskii,
{\em Statistical Physics}\/ V. 2, p. 145 (Nauka, Moscow, 1978).
\end{thebibliography}
\end{document}